\documentstyle[11pt]{article}
\input{psfig}
\begin{document}
\begin{titlepage}
\begin{flushright}
{OUTP-97-31-P}\\
{hep-ph/9707401}\\
\end{flushright}
\vskip 0.5 cm
\begin{center}
 {\Large{\bf Unification predictions}}\\
        \vskip 0.6 cm
        {\large{\bf Dumitru Ghilencea$^{*}$\footnote{
                E-mail address: D.Ghilencea1@physics.oxford.ac.uk}
                , 
                Marco Lanzagorta$^{\dagger}$\footnote{
                E-mail address: unisol@ictp.trieste.it}
                ,
                Graham G. Ross$^{*}$\footnote{
                E-mail: Ross@thphys.oxford.ac.uk}}}\\
        \vskip 1 cm
{$^{*}${\it Department of Physics, Theoretical Physics, University of Oxford}}\\
{\it 1 Keble Road, Oxford OX1 3NP, United Kingdom}\\
\vskip 0.2 cm

{$^{\dagger}$ {\it High Energy Section, International Centre for 
                                        Theoretical Physics}}\\
             {{\it PO Box 586, Trieste, Italy}}\\ 
\end{center}
\vskip 2 cm
\begin{center}
{{\it Published in Nuclear Physics {\bf B 511(1,2)}, 1998, 3-24}}
\end{center}\vskip 2 cm
\begin{abstract}
{The unification of gauge couplings suggests that there is an underlying
(supersymmetric) unification of the strong, electromagnetic and weak
interactions. The prediction of the unification scale may be the first
quantitative indication that this unification may extend to unification with
gravity. We make a precise determination of these predictions for a class of
models which extend the multiplet structure of the Minimal Supersymmetric
Standard Model to include the heavy states expected in many Grand Unified
and/or superstring theories. We show that there is a strong cancellation
between the 2-loop and threshold effects. As a result the net effect is
smaller than previously thought, giving a small increase in both the
unification scale and the value of the strong coupling at low energies.}
\end{abstract}
\end{titlepage}

\setcounter{footnote}{0}
\section{Introduction}

\noindent While the Standard Model continues to provide a remarkably
consistent description of essentially all observed phenomena, many
physicists think that there must be a more fundamental theory capable of
determining many of its features and parameters. To date, the only
quantitatively successful predictions for physics beyond the Standard Model
are the ``unification'' predictions for its gauge and Yukawa couplings and
for this reason a great deal of attention has been paid to them. At best,
the present success of these predictions gives some circumstantial evidence
for a stage of supersymmetric unification. As the measurement of the
couplings of the Standard Model improves, it is possible to test such
unification predictions with ever higher precision and the hope is this will
strengthen the case for some underlying unification of the strong,
electromagnetic and weak interactions and shed light on its specific form.
Furthermore, the scale of unification may provide us with the first
indication of unification with gravity because it is {\it predicted} in
string theories with a value close to that needed for unification. It is the
purpose of this paper to test these predictions in detail for a class of
models which allow for the addition of heavy states as expected in
essentially all Grand Unified theories and/or superstring theories.

The ``unification'' predictions for gauge couplings rest on two premises.
The first is that the $SU(3),\;SU(2)$ and $U(1)$ gauge couplings are related
at the (large) unification scale and the second is that below this scale
there is a specific multiplet content. The first premise follows if there is
a stage of Grand Unification ($SU(5),\;SO(10),\;E_{6}$ etc). However Grand
Unification is not essential. In a (compactified) string theory the gauge
couplings are related even if the gauge group is not unified. In this case
the relationship between the couplings depends on the Kac Moody level used
to construct the theory. In the simplest level-1
theory, the original string constructions yielded the ``standard''
SU(5) relationship between the gauge couplings even though the
standard gauge group is just that of Standard
Model\footnote{Subsequently, 
alternative constructions were developped
which can lead to non-standard U(1) normalisation even in level-1
theory - for example see discussion in ref. \cite{dienes1}}.
Furthermore, in
the case of the string theory the scale of gauge coupling unification is
determined in terms of the only parameter of the theory, the string tension.
Fixing this to obtain the correct gravitational strength coupling, one finds
that the unification scale should be close to the Planck scale, {\it i.e} of 
$O(10^{18}GeV)$. The fact that the unification of the gauge couplings of the
Standard Model requires a very large scale of unification of $O((1-3)\times
10^{16}) GeV$ may be considered to be the first indication that we are on
the right track when seeking to extend unification to include gravity as in
the string theory. However, one should note that the prediction of the scale
of unification of the Standard Model gauge group couplings applies only to
the class of string theory which do {\it not} have a stage of Grand
Unification.

The second premise needed for the unification predictions is the choice of
the low-energy spectrum below the unification scale. The usual ``minimal''
assumption is that this is just the minimal supermultiplet spectrum of the
minimal supersymmetric Standard Model (MSSM). However this is {\it not} the
most reasonable expectation.

In any viable Grand Unified theory it is necessary to add additional
supermultiplets to obtain the required pattern of symmetry breaking. These
additional multiplets are vectorlike with respect to the Standard Model
group, so that they may obtain very large masses of the order of the (Grand)
unification scale. However, only if they are exactly degenerate with the
additional massive gauge bosons of the Grand Unified theory, will they leave
the ``minimal'' unification predictions unchanged.

In string unification, even without Grand Unification, there is a definite
multiplet content. We noted above that even if the gauge group is just that
of that Standard Model, provided it is a level-1 string theory, the gauge
couplings may still be related at the unification scale as in the
prototype $SU(5)$ GUT. In this case, the breaking of the symmetry is 
through Wilson lines and one may show that the families of quarks and 
leptons belong to
complete $SU(5)$ multiplets even though there is no residual $SU(5)$ gauge
symmetry \cite{witold}; this is just as is required by observation. In
addition, specific Wilson line breaking allows for an additional light pair
of Higgs doublets as required in the MSSM. However, like the conventional
GUT, such string theories usually have additional states, vectorlike under
the Standard Model group, which can acquire masses of the order of the
unification scale.
In cases with a single Abelian Wilson line associated  with a freely
acting discrete symmetry, these states must fill out complete $SU(5)$
representations or come as additional copies of the Higgs sector; no other
representations are allowed even though there is no Grand Unified group. It
is worth remarking that the presence of complete degenerate $SU(5)$
multiplets does not affect the unification of gauge couplings at one loop
and thus, the success of gauge unification applies in this case too.
However, additional massive Higgs doublets do spoil the unification
predictions at one loop and are disfavoured.

With the minimal spectrum of the MSSM and assuming the $SU(5)$ normalization
of the $U(1)$ gauge couplings, one may determine the three gauge couplings in
terms of the value of the unified coupling and the unification scale giving
one overall prediction (up to uncertainties due to our lack of knowledge of
the low energy SUSY spectrum). It is most convenient to express this as a
prediction for the strong coupling, as this is the least well known of the
three couplings. This gives  $\alpha _{3}(M_{z})\geq 0.126$ \cite{shifman1}
including the supersymmetric threshold corrections \cite{alfas}, using as an
input the well measured value of $\sin ^{2}\theta _{W}$ and $\alpha _{EM}$.
This is in disagreement with the world average value of $\alpha
_{3}(M_{z})=0.118\pm 0.003$ \cite{data}, albeit marginally so. In string
theories, as we have already mentioned, the gauge unification scale is also
predicted. In the case of the (weakly coupled) heterotic string the
prediction is \cite{scale}

\begin{equation}
M_{string}\approx g_{string}\times (5.2\times 10^{17}GeV)\approx 3.6\times
10^{17}GeV  \label{su}
\end{equation}
more than an order of magnitude larger than that predicted by the
MSSM. Ways to resolve this discrepancy have been recently reviewed
by Dienes \cite{dienes2}. Of particular interest is a recent
suggestion of Witten \cite{witten} that the (10 dimensional) strongly 
coupled heterotic string theory (M theory) gives a gauge coupling more
closely in agreement with the gauge unification value.

It is clear that the precision measurements of the gauge couplings already
allow for a very detailed test of the unification predictions and offer the
potential of distinguishing between different unified theories. Given this,
we consider it important to explore in detail the implications for the
unification predictions of a non-minimal spectrum at the unification scale
of the type to be expected in realistic extensions of the Standard Model. As
we have mentioned, there are likely to be additional vectorlike states with
masses of order of the unification scale belonging to complete $SU(5)$
multiplets or copies of the MSSM Higgs supermultiplets. Only if these states
have exactly the unification scale mass, will the usual MSSM predictions
apply, but the difference in the predictions if this is not the case applies
in two loop order and may be expected to be small. We will consider in
detail the effects of ($\overline{5}+5)$ and ($\overline{10}+10)$ additional 
$SU(5)$ supermultiplets which have masses less than the unification scale.
While this does not exhaust the possibilities in Grand Unified theories, as
there may be further higher dimension representations, the method we develop
to obtain precision solutions to the renormalization group equations can
readily be generalized to such representations. 
However, in the class of string theories discussed above,
the only states allowed are indeed just those that fill out the
five and ten dimensional representations, so the analysis presented here is
quite general \footnote{
We discard the possibility of additional massive Higgs representations, for
they are known to spoil the unification predictions at one loop order.}.
Furthermore, it is only in the context of such string theories that the
prediction for the string unification scale may be directly compared with
the unification scale of the couplings of the Standard Model.

The organization of the paper is as follows. In the next Section we discuss
the mass splitting of the additional massive states. This is important
because it gives rise to threshold corrections to the gauge couplings. In
the case of Grand Unification, the splitting comes from calculable radiative
corrections and the resulting threshold corrections are of the same order as
the usual two loop corrections. In the case of string unification without
Grand Unification, the components of an $SU(5)$ multiplet are not related by
the low energy symmetry and there may be additional threshold effects.

In Section 3 we develop a method for determining the low energy coupling $%
\alpha _{3}(M_{z})$ and the unification scale, correct to two loop order in
the running coupling, ignoring the effect of Yukawa couplings. The simplest
method is just to integrate numerically the coupled differential equations,
technique used in \cite{pati} and in \cite{hemp}. 
However, we found problems in achieving the
required accuracy using this technique, so we developed a method in which
the calculation is done analytically. This has the advantage of more clearly
exhibiting the important parameters in the calculation. In Section 4 we
compare our intermediate results with the form presented by Shifman \cite
{shifman}, derived by a completely different method, and show that they lead
to same answer. In Section 5 we use the analysis to determine analytic
formulae for the unification scale, the strong coupling constant and the
intermediate scale masses. Using these formulae we determine the numerical
predictions and present them in Section 6. The effect of additional
corrections to the mass of the heavy $SU(5)$ multiplets is considered in
Section 7. Finally in Section 8 we present a summary of our results and our
conclusions.

\section{The Massive spectrum}

We are interested in the class of models with additional vectorlike states $%
I+\bar{I}$ where $I$, $\bar{I}$ denote complete representations of $SU(5)$.
The components of complete $SU(5)$ representations transform under SU(3) and
SU(2) groups as follows: for $\psi=d^c :(\bar{3},1)\ $, $\psi=l: (1,2)$ for
the case $I$ is the 5 dimensional representation of $SU(5)$ and $\psi=e^c:
(1,1)\ , \psi=u^c: (\bar 3,1)\ , \psi=q: (3,2)\ $ for the case $I$ is the 10
dimensional representation of $SU(5)$. We note that in string
compactifications there are just such additional states left massless after
compactification \cite{witold}. Here we wish to discuss how they obtain a
(large) mass. Since they come in vectorlike combinations, we can write down $%
SU(3)\otimes SU(2)\otimes U(1)$ invariant mass terms $\mu_{\psi}\psi\bar{\psi%
}$ where the mass $\mu_{\psi}$ can be far above the electroweak breaking
scale. This term will arise from a stage of spontaneous symmetry breaking
through the coupling $\lambda_{\psi}\phi_{\psi}\psi\bar{\psi}$, when the $%
SU(3)\otimes SU(2\otimes U(1)$ invariant scalar field $\phi _{\psi}$
acquires a vacuum expectation value (vev). There is a very natural
explanation for the origin of this vev, because the mass squared of the
fields $\phi _{\psi}$ will be driven negative by the radiative corrections
involving the coupling $\lambda _{\psi}$, in the usual radiative breaking
mechanism \cite{IR}.

To discuss this possibility in more detail, we consider first the case of a
single $\phi $ with Yukawa coupling $\lambda_{\psi }\phi \psi 
\overline{\psi }$. The renormalization group equations have the form 
\begin{equation}
\frac{d\alpha _{i}}{dt}=b_{i}^{\prime }\alpha _{i}^{2}
\end{equation}
\label{gc}
\begin{equation}
\frac{d\ln Y_{\psi }}{dt}=-2\sum_{j=1}^{3}\eta _{j,\psi }\alpha
_{j}+2Y_{\psi }+\frac{n_{5}}{2}\left( 3Y_{d^{c}}+2Y_{l}\right) +\frac{n_{10}%
}{2}\left( 6Y_{q}+3Y_{u^{c}}+Y_{e^{c}}\right)   \label{yrg}
\end{equation}
\begin{equation}
\label{mphi}
\frac{dm_{\phi }^{2}}{dt}=\sum_{\psi }{s_{\psi }}Y_{\psi }\left( m_{\phi
}^{2}+m_{\psi }^{2}+\overline{m}_{\psi }^{2}\right)   
\end{equation}
\begin{equation}
\label{rge1n}
\frac{dm_{\psi }^{2}}{dt}=-2\sum_{j=1}^{3}\eta _{j,\psi }M_{j}^{2}\alpha
_{j}+Y_{\psi }\left( m_{\phi }^{2}+m_{\psi }^{2}+\overline{m}_{\psi
}^{2}\right)   
\end{equation}
Here $\alpha _{i}=g_{i}^{2}/{4\pi }\ $, $i=\{1,2,3\}\,$, $Y_{\psi }=\lambda
_{\psi }^{2}/(4\pi )\,$, $b_{i}^{\prime }=b_i+(n_5+3 n_{10})/2\,$ where
$b_i$ is the MSSM one loop beta function, $\eta _{j,\psi }$ are given
in the Appendix and $M_{i}$ are the gaugino masses which evolve as 
$M_{i}(t)=M_{i}(0)(\alpha _{i}(t)/\alpha _{i}(0))$. Also $m_{\phi }$, 
$m_{\psi }$ and $\overline{m}_{\psi }$ refer to the soft supersymmetry
breaking masses of the fields $\phi ,$ $\psi $ and $\overline{\psi}$
respectively\footnote{We use the notation  $n_5=N_5+N_{\bar 5}$, 
$n_{10}=N_{10}+N_{\overline 10}$}. 
The above sum over $\psi $ runs over the component fields $\psi
=\{l,d^{c},e^{c},u^{c},q\}$ for the case one considers both $5$ and $10$
dimensional complete representations of $SU(5)$ with 
$s_{\psi}=\frac{1}{2}\{2n_5, 3 n_5, n_{10},3n_{10}, 6n_{10}\}$.

Let us discuss how the field $\phi $ actually acquires a vev, and what the
expectation for its magnitude is. This requires us to solve the
renormalization group equations for the soft SUSY breaking masses $m_{\phi
}\,$, $m_{\psi }\,$, $\overline m_{\psi }$. It is straightforward to
solve these equations numerically in specific cases. To determine the
general feature, we solve a more simple example, keeping only the QCD gauge
coupling and considering the case of a {\it single} component field 
$\psi$ which  has QCD quantum numbers. 
Solving eqs (\ref{mphi}),(\ref{rge1n}) leads to the result 
$m_{\psi }=\overline{m}_{\psi}$ and

\[
m_{\phi }^{2}(t)=\left[ \frac{\alpha (t)}{\alpha(0)}
\right]^{\gamma_{m}}
\left\{m_{\phi}^2(0)+\frac{2 s_{\psi}Y_{\psi}^{*}}
{b'_3(\gamma_{\psi}-\gamma_{m})}
\left[m_{\psi}^{2}(0)+\frac{2 \eta_{j,\psi}M_{3}^{2}(0)}{b'_3(2-\gamma_{\psi})}
\right]\left[\left(\frac{\alpha(t)}{\alpha (0)}\right)
^{\gamma_{\psi}-\gamma_{m}}-1\right] \right. 
\]
\begin{equation}
\left.
-\frac{4 s_{\psi} Y_{\psi}^{*}\eta_{j,\psi}M_{3}^{2}(0)}
{b^{\prime 2}_3(2-\gamma_{m})(2-\gamma_{\psi})}
\left[\left(\frac{\alpha(t)}{\alpha (0)}
\right)^{2-\gamma_{m}}-1\right]\right\}   
\label{pn1}
\end{equation}
where $t=\frac{1}{2\pi}\ln (\mu /M_{g})$ and 
$\gamma_{m}=s_{\psi}Y_{\psi}^{*}/b'_3$, 
$\gamma_{\psi}=2 Y_{\psi}^{*}/b'_3$ where $Y_{\psi}^{*}$ denotes the fixed
point value for the ratio $Y_{\psi}/\alpha_3$ which can be easily computed
from eqs.(\ref{gc}) and (\ref{yrg}). 
Note that this equation implies that $m_{\phi}^{2}$ will be driven 
{\it negative} by the radiative corrections, triggering a vacuum 
expectation value (vev) for $\phi $. This is the general behavior
for our simplified example with one field $\psi$ only.

In the case that $\phi $ has a ``flat'' potential
\footnote{{\it i.e.} there are{\it \ no} terms in 
the potential involving $\phi.$}, the scale $\Lambda$ at which 
this happens, corresponds closely to the vev that is induced for $\phi $. 
This then gives a mass to the vectorlike fields $I$, $\bar{I}$ coupled
to $\phi$. The important point to note is that $\Lambda $ is
determined by the multiplet content and by the allowed couplings of
the theory; it is not a free parameter. 
In the case we are considering, it is easy to determine the order of 
magnitude of $\Lambda $ using that $\gamma _{\psi}\propto 1/n_{10}$ 
(or $1/n_5$) and $\gamma_{m}\sim 1$. Hence, for large $n_{10}\,(n_5)$
case, appropriate to our use of the fixed points in solving these 
equations,  and with $M_{3}(0)/m_{\psi }(0)<<1$, we get from
eq. (\ref{pn1}) that:  
\begin{equation}
m_{\phi }^{2}(t)=\left[ \frac{\alpha (t)}{\alpha (0)}\right] ^{\gamma _{m
}}\left\{ m^2_{\phi }(0)-2 m^2_{\psi}(0)\left[\left( \frac{\alpha (t)}{\alpha (0)}\right) ^{-\gamma_{m}}-1\right]\right\}   \label{pn}
\end{equation}
 The point at
which $m_{\phi}^{2}$ is driven negative is clearly dependent on the initial
value of $m_{\phi}/m_{\psi}$. With this of order 1, it will occur at the
point where $\alpha (t)/\alpha (0)$ deviates from 1 by $O(1)$. For the
case of large number of extra multiplets, this will be very close to the 
unification scale $M_{g}$, so $\Lambda $ will be close to this scale too.

Although, as we have just shown, it is possible that the intermediate scale
breaking is triggered close to the unification scale, for $\alpha
(t)/\alpha (0)$ of $O(1)$, it is also possible for a much smaller
intermediate scale to be generated, corresponding to a much smaller ratio of 
$\alpha (t)/\alpha (0)$, its magnitude depending sensitively on this ratio.
	
Even if these parameters are of the same order, the intermediate scale
may be much smaller than our initial estimate. If $\phi $ does not 
have a ``flat'' potential, then the scale of the vev of $\phi$ is not 
determined by the scale at which its mass squared becomes negative. 
To see this, let us assume the potential with leading term 
$\phi ^{r}/M^{r-4}$ 
\footnote{The value of $r$ will be determined by the symmetries of the
theory.} , where $M$ is the scale of the new physics leading to the effective potential,
which we take to be the compactification or Planck scale. Then, the
vev of $\phi $ is proportional to $(m_{\phi }^{2}M^{r-4})^{1/(r-2)}$. 
Since the SUSY breaking scale $m_{\phi}$ is of $O(1TeV)$, this implies
an intermediate scale ranging from $10^{10}GeV$ to $10^{15}GeV$ for 
$r$ between 6 and 12.

To summarize, there is a very natural origin for the generation of the
intermediate scale, coming from radiative breaking. The magnitude of the
intermediate scale is sensitively dependent on the details of the theory.

Let us now consider the expectation for individual intermediate scale
masses. 

For the case of Grand Unification, $SU(5)$ or larger, the mass terms
must be $SU(5)$ invariant, so there is only a single field $\phi $ and $%
\lambda _{\psi }$ is independent of $\psi $. Thus, at the unification scale,
all components $\psi $ of a representation $I$ are degenerate. However,
below this scale, until the scale $M_{I}=<\phi >$, radiative corrections
will split the mass of the components. For the case when one can ignore the
Yukawa couplings such as $\lambda _{\psi }$ in computing the radiative
corrections, this splitting is generated by the gauge interactions only and
is thus easily calculable. The effects of this splitting is included in the
determination of the gauge unification predictions given in the next
Section. The case where Yukawa couplings are not negligible is discussed
below and in Section 7.

For the case of string unification, even without Grand Unification, the
gauge couplings will be unified. Indeed, if the string prediction for the
unification scale is relevant, there {\it must not} be a stage of Grand
Unification because the prediction refers to the unification scale for the
couplings of the gauge group at compactification scale, and this must just
be that of the Standard Model. However, without Grand Unification, there may
be different fields $\phi _{\psi }$ responsible for giving mass to the
different components of the $SU(5)$ representation. Furthermore, there is no
reason why the couplings $\lambda _{\psi }$ should be the same for all $\psi 
$'s. 

Let us first consider the case that there is only one field $\phi $. In this
case the Yukawa coupling  $\lambda _{\psi }\phi \psi \overline{\psi }$ leads
to a mass $\lambda _{\psi }<\phi >$ where $<\phi >$ is the vev of $\phi .$
The individual couplings $\lambda _{\psi }$ get renormalised according to eq(%
\ref{yrg}) causing the spectrum to split. Just as for the $SU(5)$
case, the gauge contributions (the first term on the rhs) are included
in our analysis below. The case the Yukawa couplings are large  
is discussed in Section 7. This leaves only the corrections which
occur if the bare couplings $\lambda_{\psi }$ are different and the 
effect of this is also presented in Section 7.

Finally we consider the  possibility  when there are different
fields $\phi _{\psi }$ responsible for the masses of the various
components $\psi=\{l,d^{c},e^{c},u^{c},q\}$ of the representations of 
$SU(5)$. Again radiative effects will change the couplings 
$\lambda _{\psi }.$ However, there is a more important effect, 
because the radiative breaking mechanism
giving rise to the $\phi _{\psi }$ vev, drives $m_{\phi _{\psi }}$ negative
at different scales, generating different vevs. The terms responsible for
this in the renormalization group equations are those proportional to the
gauge coupling strength of the component field $\psi $. The net effect of
this is that the $(3,1)$ components of the 5 dimensional are heavier than
the $(1,2)$ components, and the components in the 10 dimensional
representation are similarly ordered with $(3,2)>(3,1)>(1,2)>(1,1)$. This
effect is in the same direction as that driven by the gauge radiative
corrections to the couplings alone, but generates a larger splitting between
the components. Again the effect of this may be determined using the results
of Section 7.

\section{Renormalization group equations}

We consider the two loop renormalization group equations for the gauge
couplings, with no top, bottom and tau Yukawa interaction. Their
general form is given by: 
\begin{equation}
\frac{d\alpha _{i}}{dt}={\tilde{b}_{i}}\alpha _{i}^{2}+\frac{1}{4\pi }%
\sum_{j=1}^{3}{\tilde{b}_{ij}}\alpha _{i}^{2}\alpha _{j}+{\cal O}(\alpha
^{4})  \label{rge}
\end{equation}
with $i=\{1,2,3\}$ and where $t=\frac{1}{2\pi }\log {Q/M_{g}}$; $M_{g}$ is
the unification scale, and $\tilde{b}_{i}$ and $\tilde{b}_{ij}$ are, 
in a general notation, the  one loop and two loop beta functions 
respectively, which depend on the multiplet content of the theory.
Dividing both sides by $\alpha _{i}^{2}$ we may rewrite this equation in the
form 
\begin{equation}
d\left( -{\alpha _{i}}^{-1}\right) ={\tilde{b}_{i}}dt+\frac{1}{4\pi }%
\sum_{i=1}^{3}\frac{{\tilde{b}_{ij}}}{\tilde{b}_{j}}d(\ln \alpha _{j})+{\cal %
O}(\alpha ^{2})  \label{rge1}
\end{equation}
where we have used $\alpha _{j}dt=\frac{1}{\tilde{b}_{j}}\frac{d\alpha
_{j}}{\alpha _{j}}+{\cal O}(\alpha ^{2})$ to rewrite the second term on the
right hand side. We integrate this equation from $M_{g}$ scale down to
$M_{z} $ scale, and replace $\tilde b_i$ and $\tilde b_{ij}$ with their
appropriate values. As discussed in the
previous section, the mass splitting of the additional $SU(5)$ multiplets
added to the MSSM spectrum are generated by radiative corrections. Thus to $%
{\cal O}(\alpha ^{2})$, the threshold effects coming from this splitting
need only be included at one loop order; the two loop terms may be cutoff at
the mean mass, $\mu _{q}$ of the multiplet. This gives 
\begin{equation}
\alpha_i^{-1}(M_z)=-\delta_i+\alpha_g^{-1}+\frac{b_i^{\prime}}{2\pi}\ln
\left[\frac{M_g}{M_z}\right]+\zeta_i +\frac{1}{4\pi}\sum_{j=1}^{3}\frac{%
b^{\prime}_{ij}}{b^{\prime}_j}\ln\left[\frac
{\alpha_g}{\alpha_j(\mu_q)}\right]+\frac{1}{4\pi}\sum_{j=1}^{3} \frac{b_{ij}%
}{b_j}\ln\left[\frac{\alpha_j(\mu_q)}{\alpha_j(M_z)}\right] 
\label{coupling}
\end{equation}

\noindent with 
\begin{equation}  \label{ren}
\zeta_i=-\frac{1}{2\pi}\sum_{\psi}{}N_{\psi}\beta_{i,\psi} \ln\left[\frac{%
\mu_\psi(\mu_\psi)}{M_z}\right]
\end{equation}
In the above equations $\delta_i$ stands for low energy supersymmetric
thresholds; we will express our results as a change of the MSSM predictions
following from the addition of heavy states and in this the $\delta_i$
cancel; further,  $\psi$ stands for the component fields of additional
complete $SU(5)$ multiplets we consider and runs over
$\{l,d^c,e^c,u^c,q\}$; their mass is $\mu_{\psi}$; 
$b^{\prime}_{i}$ and $b^{\prime}_{ij}$ are the one loop and two
loop beta functions when none of these states is decoupled\footnote{%
We assume here that all additional states of a given flavour are degenerate;
it is easy to generalise the method to the non-degenerate case if desired.};
their expressions are presented in the Appendix. The above formula is
correct to two loop order.

As we have stressed, the one loop threshold effects $\zeta_i$ are in fact of
the same order as the two loop terms. This is made explicit from the one
loop renormalization group equations for the running mass 
\begin{equation}  \label{psi}
\frac{d\mu_{\psi}}{dt}=-\mu_{\psi}\sum_{j=1}^{3}\alpha_j\eta_{j,\psi}
\label{olr}
\end{equation}
where $\eta_{j,\psi}$ are given in the Appendix and we have ignored Yukawa
interactions involving the $\psi$. Keeping terms to ${\cal O}(\alpha^2)$
corresponds to keeping only the one loop running approximation for the
couplings, in the presence of extra-matter: 
\begin{equation}  \label{onel}
\frac{d\alpha_i}{dt}=b^{\prime}_i \alpha_i^2+{\cal O}(\alpha^3)
\end{equation}
Using this we may integrate eq(\ref{olr}) to get
\begin{equation}  \label{mupsi}
\ln\left[\frac{\mu_{\psi}(\mu_{\psi})}{\mu_g}\right]=\sum_{j=1}^{3} \frac{%
\eta_{j,\psi}}{b^{\prime}_j} \ln\left[\frac{\alpha_g}{\alpha_j(\mu_{\psi})}%
\right]
\end{equation}
where we have assumed that all the heavy states get a common mass $\mu_g$ at
unification scale. One sees explicitly that one loop threshold effects are,
in fact, two loop-like effects. As before we may further simplify this
equation to write 
\begin{equation}  \label{muq}
\ln\left[\frac{\mu_{\psi}(\mu_{\psi})}{\mu_g}\right]=\sum_{j=1}^{3} \frac{%
\eta_{j,\psi}}{b^{\prime}_j} \ln\left[\frac{\alpha_g}{\alpha_j(\mu_{q})}%
\right]
\end{equation}
for any $\psi$, since the effect of the radiative mass difference between
different flavours generates terms of higher order.

Using eq. (\ref{muq}) to express  $\zeta_i$  given in eq.(\ref{ren}), 
we get from eq.(\ref{coupling}) the value of the  couplings at $M_z$ scale:

\begin{eqnarray}  \label{HMSSM}
&\alpha_i^{-1}(M_z)=&-\delta_i+\alpha_g^{-1}+\frac{b_i}{2\pi}\ln \left[\frac{%
M_g}{M_z}\right]+\frac{n}{2\pi}\ln\left[\frac{M_g}{\mu_g} \right]-\frac{1}{%
2\pi}\sum_{j=1}^{3}Y_{ij}\ln\left[\frac {\alpha_g}{\alpha_j(\mu_q)}\right] 
\nonumber \\
& &+\frac{1}{4\pi}\sum_{j=1}^{3} \frac{b_{ij}}{b_j}\ln\left[\frac{\alpha_g}{%
\alpha_j(M_z)}\right]
\end{eqnarray}
where 
\begin{equation}  \label{why}
Y_{ij}=\frac{n}{b_j^{\prime}}\left[\frac{1}{2}\frac{b_{ij}}{b_j}-
\delta_{ij}\lambda_j\right]
\end{equation}
and $\lambda_1=0$, $\lambda_2=2$, $\lambda_3=3$.

These equations demonstrate an interesting result. At one loop the running
of the couplings depends only on the number of additional states through the
combination $n=(n_5+3n_{10})/2$. We see that is also true at two loop, even
though it is not true for the two loop beta functions, because of the
threshold effects. Note that this only applies in the case Yukawa couplings
are ignored. Note also the mild dependence of the functions $Y_{ij}$ on $n$.

In order to evaluate the effect of the additional heavy states, it is
convenient to compare the results with heavy states to the minimal MSSM
predictions. In the latter, the Renormalization Group Equations (RGE) are 
\begin{equation}
\alpha _{i}^{o-1}(M_{z})=-\delta _{i}^{o}+\alpha _{g}^{o-1}+\frac{b_{i}}{%
2\pi }\ln \left[ \frac{M_{g}^{o}}{M_{z}}\right] +\frac{1}{4\pi }%
\sum_{j=1}^{3}\frac{b_{ij}}{b_{j}}\ln \left[ \frac{\alpha _{g}^{o}}{\alpha
_{j}^{o}(M_{z})}\right]  \label{MSSM}
\end{equation}
again ignoring Yukawa couplings. The MSSM variables are labelled with an
``o'' index to distinguish them from the model which considers additional
complete SU(5) multiplets. We follow the general approach and fix, as an
input, the values for $\alpha _{1}(M_{z})$ and $\alpha _{2}(M_{z})$ which
are known with a good accuracy from the experimental values for $\alpha
_{em}(M_{z})$ and $\sin \theta _{W}(M_{z})$. This fixes the values for $%
\alpha _{g}^{o}$ and $M_{g}^{o}$ which are then used to compute numerically
the value for $\alpha _{3}^{o}(M_{z})$. As the MSSM has been extensively
studied numerically, we can take all these MSSM quantities as known.

Note that the parameters $\delta _{i}^{o}$ which appear in eq(\ref{MSSM})
and the parameters $\delta _{i}$ which appear in eq(\ref{ren}) include the
constant factors needed for ${\overline{MS}}\rightarrow {\overline{DR}}$
(which are the same in MSSM and our model \footnote{%
and will therefore cancel in the calculation below}), the low energy
supersymmetry thresholds effects and contributions from nonrenormalizable
operators at the high scale \cite{lan}.  The SUSY threshold effects are most
conveniently included in terms of the effective scale $T_{susy}^{o}$ \cite
{tsusy} relevant when calculating the strong coupling in terms of the weak
and electromagnetic couplings. Thus, when comparing the prediction with
heavy states to the MSSM prediction we find 
\begin{equation}
\alpha _{3}^{-1}(M_{z})=\alpha _{3}^{o-1}(M_{z})+\Delta \alpha
_{3}^{-1}(M_{z})
\end{equation}
where the SUSY threshold effects give a contribution 
\[
\Delta \alpha _{3}^{-1}(M_{z})=\frac{19}{28\pi }\ln \left[ \frac{T_{susy}}{%
T_{susy}^{o}}\right] 
\]
and $T^{susy}$ is the new effective scale in MSSM plus the extra heavy
states. At one-loop order, the SUSY threshold factors are independent of the
gauge couplings and thus are the same for the MSSM and the MSSM with
additional heavy states. At two loop order, the SUSY threshold factors do
depend on gauge couplings, but the difference between these couplings in the
models with heavy states is $O(\alpha )$ and thus the net effect is beyond
the two-loop order and can be neglected. The same arguments enable us to
avoid the explicit form of SUSY thresholds when computing other variables of
the model with heavy multiplets, like $M_{g}$ or $\mu _{\psi }(\mu _{\psi })$.

\section{Compatibility with other approaches.}

After completing this analysis our attention was brought to the analysis of
Shifman \cite{shifman} derived from a different point of view using
holomorphicity arguments and the instanton calculus. Here we wish to compare
the results of the two approaches, namely the comparison of eq.(\ref{HMSSM})
with Shifman's ``master formula'' \cite{shifman}. We reproduce it below
(ignoring the heavy Higgs triplet and 24-plet contributions \footnote{%
These are not included in this work.}, but including the other heavy states'
contributions; we also add some generic values, $\delta _{i}$, for the low
energy Susy thresholds, as we want the values of couplings at $M_{z}$ scale.
We have: 
\begin{eqnarray}
\alpha _{1}^{-1}(M_{z}) &=&-\delta _{1}+\alpha _{g}^{-1}+\frac{1}{2\pi }%
\frac{3}{10}\left\{ \sum_{gen}{}\left[ \ln \frac{M_{g}}{M_{z}Z_{lL}}+2\ln 
\frac{M_{g}}{M_{z}Z_{eR}}+\frac{1}{3}\ln \frac{M_{g}}{M_{z}Z_{qL}}\right.
\right.   \nonumber \\
&&  \nonumber \\
&&+\left. \left. \frac{8}{3}\ln \frac{M_{g}}{M_{z}Z_{uR}}+\frac{2}{3}\ln 
\frac{M_{g}}{M_{z}Z_{dR}}\right] +2\ln
\frac{M_{g}}{M_{z}Z_{H_{u,d}}}\right\}+
\frac{n}{2\pi }\ln \frac{M_{g}}{\mu _{g}}
\end{eqnarray}

\begin{eqnarray}
\alpha_2^{-1}(M_z)&=&-\delta_2+\alpha_g^{-1}-\frac{6}{2\pi}\ln\frac{M_g}{M_z
\left(\frac{\alpha_g}{\alpha_2(M_z)}\right)^{1/3}} +\frac{1}{2\pi}%
\sum_{gen}{}\left[\frac{3}{2}\ln\frac{M_g}{M_z Z_{qL}} +\frac{1}{2}\ln\frac{%
M_g}{M_z Z_{lL}}\right]  \nonumber \\
\nonumber \\
& & +\frac{1}{2\pi}\ln\frac{M_g}{M_z Z_{H_{u,d}}} +\frac{n}{2\pi}
\ln\frac{M_g}{\mu_g}
\end{eqnarray}

\begin{eqnarray}
\alpha _{3}^{-1}(M_{z}) &=&-\delta _{3}+\alpha _{g}^{-1}-\frac{9}{2\pi }\ln 
\frac{M_{g}}{M_{z}\left( \frac{\alpha _{g}}{\alpha _{3}(M_{z})}\right) ^{1/3}%
}+\frac{1}{2\pi }\sum_{gen}{}\left[ \ln \frac{M_{g}}{M_{z}Z_{qL}}+\frac{1}{2}%
\ln \frac{M_{g}}{M_{z}Z_{uR}}\right.\nonumber\\
&&  \nonumber \\
&&+\left. \frac{1}{2}\ln \frac{M_{g}}{M_{z}Z_{dR}}\right] +\frac{n}{2\pi }%
\ln \frac{M_{g}}{\mu _{g}}
\label{shif}
\end{eqnarray}
The advantage of this form for the running of gauge couplings is that (above
supersymmetric scale, when $\delta _{i}=0$) this form is {\it exact} to all
orders. However the wave function renormalization coefficients $Z$ are only
known perturbatively so one is still confined to a given order in
perturbation theory when testing coupling unification. To two loop order,
one must include the values of wave-function renormalization coefficients $Z$
in one loop only; also, replacing them with their two loop value would give
a three loop calculation and so on \footnote{%
Beyond two loop order this form is regularisation scheme dependent \cite
{jones}. }. Here we only show that, using their one loop values, we get the
same {\it two loop} form for the running of gauge couplings as we got in
equation (\ref{HMSSM}). The $Z$ coefficients are, to one loop, given by: 
\begin{equation}
Z_{SU(j)}=\left[ \frac{\alpha _{g}}{\alpha _{j}(\mu _{q})}\right] ^{-\frac{%
2C_{2}}{b_{j}^{\prime }}}\left[ \frac{\alpha _{j}(\mu _{q})}{\alpha
_{j}(M_{Z})}\right] ^{-\frac{2C_{2}}{b_{j}}}=\left[ \frac{\alpha _{g}}{%
\alpha _{j}(\mu _{q})}\right] ^{\frac{2C_{2}}{b_{j}^{\prime }}\frac{n}{b_{j}}%
}\left[ \frac{\alpha _{g}}{\alpha _{j}(M_{Z})}\right] ^{-\frac{2C_{2}}{b_{j}}%
}
\end{equation}
where $j=1,2,3$, $\,C_{2}$ is the second order Casimir operator \footnote{%
divided by $Y^2$ for U(1), hence giving a 3/5 factor, with $Y$ defined as in
the Appendix; our definition for $Y$ differs from that of \cite{shifman} and
hence the different powers for $Z_{U(1)}$ in the following equations} for $%
SU(N)$, and $b_{j}^{\prime }=b_{j}+n$ and where we took into consideration
the one loop renormalization of these coefficients.
Note that we can neglect  the splitting of the heavy $SU(5)$ multiplets in the
renormalization of the $Z$ factors as they correspond to higher than
second order corrections in the RGE for the couplings\footnote{
Strictly speaking one should use the bare mass $\mu _{g}$ in the expressions
for $Z$ instead of $\mu _{q}$, but again, this difference is of higher
order.}. We get that: 
\begin{equation}
Z_{SU(3)}=\left[ \frac{\alpha _{g}}{\alpha _{3}(\mu _{q})}\right] ^{-\frac{8
}{9}\frac{n}{b_{3}^{\prime }}}\left[ \frac{\alpha _{g}}{\alpha _{3}(M_{Z})}
\right] ^{\frac{8}{9}}
\end{equation}
\begin{equation}
Z_{SU(2)}=\left[ \frac{\alpha _{g}}{\alpha _{2}(\mu _{q})}\right]
^{\frac{3}{2}\frac{n}{b_{2}^{\prime }}}\left[ \frac{\alpha _{g}}
{\alpha _{2}(M_{Z})}\right] ^{-\frac{3}{2}}
\end{equation}
\begin{equation}
Z_{U(1)}=\left[ \frac{\alpha _{g}}{\alpha _{1}(\mu _{q})}\right]
^{\frac{2}
{11}\frac{n}{b_{1}^{\prime }}}\left[ \frac{\alpha _{g}}{\alpha _{1}
(M_{Z})}\right] ^{-\frac{2}{11}}
\end{equation}
Also we have that
\[
Z_{qL}=Z_{SU(3)}\times Z_{SU(2)}\times Z_{U(1)}^{1/36}
\]
\[
Z_{uR}=Z_{SU(3)}\times Z_{U(1)}^{4/9}
\]
\[
Z_{dR}=Z_{SU(3)}\times Z_{U(1)}^{1/9}
\]
\[
Z_{lL}=Z_{SU(2)}\times Z_{U(1)}^{1/4}
\]
\[
Z_{H_{u,d}}=Z_{SU(2)}\times Z_{U(1)}^{1/4}
\]
\[
Z_{eR}=Z_{U(1)}
\]
With these expressions for $Z$ coefficients we get from the ``master
formula'' that: 
\begin{eqnarray}
\alpha _{i}^{-1}(M_{z})&=&-\delta _{i}+\alpha _{g}^{-1}+\frac{b_{i}}{2\pi }%
\ln \left[ \frac{M_{g}}{M_{z}}\right] +\frac{n}{2\pi }\ln \left[ \frac{M_{g}%
}{\mu _{g}}\right] +\frac{1}{4\pi }\sum_{j=1}^{3}\frac{n}{b_{j}^{\prime }}%
\left[ 2\delta _{ij}\lambda _{j}-\frac{b_{ij}}{b_{j}}\right] \ln \left[ 
\frac{\alpha _{g}}{\alpha _{j}(\mu _{q})}\right]\nonumber\\  
\label{shifman}
&&\nonumber\\
&&+\frac{1}{4\pi }\sum_{j=1}^{3}\frac{b_{ij}}{b_{j}}\ln \left[ \frac{\alpha
_{g}}{\alpha _{j}(M_{z})}\right]
\end{eqnarray}
where the $b_{ij},\,b_{j}$ above are just the values of two loop and one
loop beta functions for MSSM. This is exactly the formula we previously
obtained in eq.(\ref{HMSSM}).

\section{Analytical Results}

We now  proceed to solve the two loop RGE eqs.(\ref{HMSSM})
analytically to obtain predictions for the model with additional
$SU(5)$ multiplets.
In both the analysis of the minimal MSSM unification and the non-minimal
case we adopt, we follow the usual analysis method of using the measured
weak and electromagnetic couplings to determine the strong coupling and the
unification scale. Thus, we fix as input parameters at the full two loop
order $\alpha_1(M_z)=\alpha_1^o(M_z)$ and $\alpha_2(M_z)=\alpha_2^o(M_z)$
and equal to their experimental value \footnote{%
The couplings are normalized such as $\alpha_1=\alpha_2=\alpha_3$ at the
unification scale.}. In the case of $n$ additional $SU(5)$ heavy multiplets,
the mean mass of these multiplets introduces a further parameter which we
choose to be $\alpha_g$. The output parameters of our model will be the
unification scale, $M_g$, the value of alpha strong at the $M_z$ scale, $%
\alpha_3(M_z)$, the masses of the additional heavy $SU(5)$ multiplets $%
\mu_{\psi}$ as well as their common mass $\mu_g$, at the unification scale.
Thus, making use of eq.(\ref{MSSM}), we solve analytically, to two loop
order 
\footnote{This means that, in the following , we can discard terms which originate
from three loop order in {\it either} MSSM or {\it in} the model with
additional multiplets.}, the system of equations (\ref{muq}), (\ref{HMSSM}),
(\ref{why}) for the unknowns $\,M_g\,$,$\,\mu_g\,$,$\,\mu_{\psi}\,$,
and $\alpha_3(Mz)$ in terms of MSSM variables $\,M_g^o\,$, $\, \alpha^{o}_3(M_z)\,
$ and $\alpha_g^o$. This will avoid any numerical evaluation of the system
of these equations, as well as the inconveniences deriving from possible
numerical instabilities.

To achieve this, we subtract eq.(\ref{MSSM}) from eq.(\ref{HMSSM}) for
same $i$ and obtain the following relation:

\[
\alpha _{i}^{-1}(M_{z})-\alpha _{i}^{o-1}(M_{z})=\alpha _{g}^{-1}-\alpha
_{g}^{o-1}+\frac{b_{i}}{2\pi }\ln \left[
\frac{M_{g}}{M_{g}^{o}}\right]+\frac{n}{2\pi}\ln
\left[\frac{M_{g}}{\mu_{g}}\right] -\frac{1}{2\pi }%
\sum_{j=1}^{3}Y_{ij}\ln \left[ \frac{\alpha _{g}}{\alpha _{j}(\mu _{q})}%
\right] 
\]
\begin{equation}
+\frac{1}{4\pi }\sum_{j=1}^{3}\frac{b_{ij}}{b_{j}}\ln \left[ \frac{\alpha
_{g}}{\alpha _{g}^{o}}\right] -\frac{1}{4\pi }\frac{b_{i3}}{b_{3}}\ln \left[ 
\frac{\alpha _{3}(M_{z})}{\alpha _{3}^{o}(M_{z})}\right]  \label{ecuatia1}
\end{equation}
Consider now the equations obtained from the above equation for $i=1$ and $%
i=2$, and subtract them to get: 
\[
0=\frac{b_{1}-b_{2}}{2\pi }\ln \left[ \frac{M_{g}}{M_{g}^{o}}\right] -\frac{1%
}{2\pi }\sum_{j=1}^{3}\left( Y_{1j}-Y_{2j}\right) \ln \left[ \frac{\alpha
_{g}}{\alpha _{j}(\mu _{q})}\right] +\frac{1}{4\pi }\sum_{j=1}^{3}\frac{%
b_{1j}-b_{2j}}{b_{j}}\ln \left[ \frac{\alpha _{g}}{\alpha _{g}^{o}}\right] 
\]
\begin{equation}
-\frac{1}{4\pi }\frac{b_{13}-b_{23}}{b_{3}}\ln \left[ \frac{\alpha
_{3}(M_{z})}{\alpha _{3}^{o}(M_{z})}\right]  \label{ecuatia2}
\end{equation}
The eq.(\ref{ecuatia2}) will give us the ratio $\frac{M_{g}}{M_{g}^{o}}$
(see later). The eq. (\ref{ecuatia1}) for $i=1$ will give us the value of
the variable $\xi =1/(2\pi )\ln (M_{g}/\mu _{q})$ which we introduce instead
of $\mu _{q}$ as an independent variable and which proves useful in solving
the system. To see this we make use of the eq.(\ref{muq}) considered for $%
\psi =q$: 
\begin{equation}
-2\pi \xi -\ln \left[ \frac{\mu _{g}}{M_{g}}\right] =\sum_{j=1}^{3}\frac{%
\eta _{j,\psi }}{b_{j}^{\prime }}\ln \left[ \frac{\alpha _{g}}{\alpha
_{j}(\mu_{q})}\right]
\end{equation}
to substitute the ratio $\frac{\mu _{g}}{M_{g}}$ in eq.(\ref{ecuatia1}).
With the above three equations we compute $M_{g}/M_{g}^{o}$, $\mu _{g}/M_{g}$
(and therefore $\mu _{g}/M_{g}^{o}$) and $\xi $ in function of the remaining
terms present in the above equations. We also observe that $\ln (\alpha
_{3}(M_{z})/\alpha _{3}^{o}(M_{z}))=0$ in two loop order - hence we can drop
it - as we can see it from the fact that $\alpha _{g}^{-1}-\alpha
_{g}^{o-1}+n\xi =0$ in one loop order. We also extensively use that $\ln
(\alpha _{g}/\alpha _{j}(\mu _{q}))=\ln (1+\alpha _{g}b_{j}^{\prime }\xi )$
which is correct to two loop order. We therefore end up with expressions for 
$M_{g}/M_{g}^{o}$, $\mu _{g}/M_{g}^{o}$ in terms of $\xi $ and with a
nonlinear equation for $\xi $ itself. On this latter equation we again
observe that $\alpha _{g}^{-1}-\alpha _{g}^{o-1}+n\xi =0$ in one loop order
as all log's present are two-loop or higher order log's. Hence the $\xi $
dependence of $M_{g}/M_{g}^{o}$, $\mu _{g}/M_{g}^{o}$ is finally lost, and
the three equations decouple, as we can always replace $\ln (1+\alpha
_{g}b_{j}^{\prime }\xi )$ by its corresponding value obtained for a $\xi $
computed in one loop.\footnote{
We will retain, however, a two loop term in the final results, to get an
idea of the importance of three loop contributions}

To solve for $\alpha_3(M_z)$ we simply multiply eq.(\ref{ecuatia1})
for $i=1,2,3$ by $b_2-b_3$, $b_3-b_1$ and $b_1-b_2$ respectively, add them
together and make use of the above observations. In all these processes the
difference $\delta_i-\delta_i^o$ is dropped as it would bring a higher order
contribution, as we mentioned in the previous section.

Finally, the remaining masses of the fields $\psi $ can be computed easily
from equation (\ref{muq}).

After this algebra and consistent neglect of higher order terms, which
proves to be essential in simplifying the equations, we get the following
analytical results:

\begin{equation}  \label{res1}
\frac{M_g}{M_g^o}=\left[\frac{\alpha_g}{\alpha_g^o} \right]^{\frac{31}{21}%
}\prod_{j=1}^{3} \left\{1+\frac{b^{\prime}_j}{n}\left[\frac{\alpha_g} {%
\alpha^{o}_g}-1-\frac{1168}{231\pi} \alpha_g\ln\left[\frac{\alpha_g}{%
\alpha_g^o}\right] \right]\right\}^{\rho_j}
\end{equation}

\begin{equation}  \label{res4}
\alpha_3^{-1} (M_z)=\alpha_3^{o -1}(M_z)-\frac{470}{77 \pi}\ln\left[\frac{%
\alpha_g}{\alpha_g^o}\right]+ \sum_{j=1}^{3}\ln\left\{1+\frac{b^{\prime}_j}{n%
} \left[\frac{\alpha_g}{\alpha_g^o}-1-\frac{1168}{231\pi} \alpha_g\ln\left[%
\frac{\alpha_g}{\alpha_g^o}\right] \right]\right\}^{\omega_j}
\end{equation}

\begin{equation}  \label{res2}
\frac{\mu_g}{M_g^o}=\left[\frac{\alpha_g}{\alpha_g^o} \right]^{\frac{31}{21}+%
\frac{2336}{231 n}} \exp\left[\frac{2 \pi}{n}\left(\alpha_g^{-1}-\alpha_g^{o
-1} \right)\right] \prod_{j=1}^{3}\left\{1+\frac{b^{\prime}_j}{n} \left[%
\frac{\alpha_g}{\alpha^o_g}-1-\frac{1168}{231\pi} \alpha_g\ln\left[\frac{%
\alpha_g}{\alpha_g^o}\right] \right]\right\}^{\sigma_j}
\end{equation}

\begin{equation}  \label{res3}
\frac{\mu_{\psi}}{M_g^o}=\frac{\mu_g}{M_g^o} \prod_{j=1}^{3}\left\{1+\frac{%
b^{\prime}_j}{n} \left[\frac{\alpha_g}{\alpha^o_g}-1-\frac{1168}{231\pi}
\alpha_g\ln\left[\frac{\alpha_g}{\alpha_g^o}\right]\right]
\right\}^{\tau_{j,\psi}}
\end{equation}
The values of $\rho_j\,$,$\,\omega_j\,$,$\,\sigma_j\,$,$\,\tau_{j,\psi}\,$,
with $j=\{1,2,3\}$ and $\psi=\{l,d^c,e^c,u^c,q\}$, are given by 
\begin{equation}
\rho_j=\left\{ \frac{1}{12}\frac{n}{b_1^{\prime}}, {\frac{-39}{28}}{\frac{n}{%
b_2^{\prime}}}, {\frac{4}{21}}{\frac{n}{b_3^{\prime}}}\right\}
\end{equation}

\begin{equation}
\omega_j=\left\{\frac{-2}{11 \pi}\frac{n}{b_1^{\prime}}, \frac{81}{14 \pi}%
\frac{n}{b_2^{\prime}}, \frac{2}{7 \pi}\frac{n}{b_3^{\prime}}\right\}
\end{equation}
\begin{equation}
\sigma_j=\left\{{\frac{11 n-7 }{132\: b_1^{\prime}}}, {\frac{-3(111+13 n)}{%
28\:b_2^{\prime}}}, {\frac{4(22+n)}{21\: b_3^{\prime}}}\right\}
\end{equation}
and finally, 
\begin{equation}  \label{res5}
\tau_{j,\psi}=\frac{\eta_{j,\psi}}{b^{\prime}_j}
\end{equation}
The values of $\eta_{j,\psi}$ and ${b^{\prime}_j}$ appearing above are given
in the Appendix.

This gives the full analytical solution, to two loop order, to the system of
non-linear equations (\ref{muq}),(\ref{HMSSM}),(\ref{why}), which contains
three RGE equations and five equations for the running of the masses of $\psi
$'s.

The term $1168/231/\pi \ln (\alpha _{g}/\alpha _{g}^{o})$ in the above
expressions can be dropped, as it brings in a higher order ({\it three loop}%
) correction; however, we keep it, as it gives an idea of the size of three
loop corrections and makes a better agreement with the numerical approach of
this problem. It is also relevant when considering unification at strong $%
\alpha _{g}$. This is equivalent to considering that $\alpha
_{g}^{-1}-\alpha _{g}^{o-1}+n\xi =-1168/231/\pi \ln (\alpha _{g}/\alpha
_{g}^{o})$ where the right hand side of the latter is a two loop correction 
\footnote{%
See Appendix for a full two loop equation}.

The solution demonstrates explicitly that the result depends on the number
of additional states only, via the parameter $n=(n_5+3n_{10})/2$. Also, the
values of $\alpha_3(M_z)$ as well as the ratio $\frac{M_g}{M_g^o}$ depend, 
{\it to two loop}, on {\it the ratio} of the couplings at unification scale
only. As a check of our calculation, one can take the limit $%
\alpha_g/\alpha_g^o\rightarrow 1$ in the above expressions to recover the
MSSM case.

\section{Numerical results}

In this section we consider the numerical results given by the expressions
of eqs. (\ref{res1}), (\ref{res4}), (\ref{res2}), (\ref{res3}).

For $n=1,2$ the ratio $M_{g}/M_{g}^{o}$ is rather small, though larger than
unity; in the meantime $\alpha _{3}(M_{z})$ is always larger than its MSSM
value; it cannot became smaller than $\alpha _{3}^{o}(M_{z})$ because
\footnote{We use $\alpha _{3}^{o}(M_{z})=0.125$}, in
that case, the additional states would be too light $(\approx 10-100GeV)$
and these cases are therefore disfavoured. Moreover in these cases the two
loop approximation breaks down for $\alpha_{g}$ quite close to its MSSM
value, because the two loop contributions involve log's of very small
arguments, which make their contribution dominate over the 
one loop contribution.

The case $n=3$ is special because the step used in deriving eq(\ref{rge1}) 
breaks down as it involves division by a one loop beta function which is
actually 0. This case should be considered separately, even though,
interesting enough, the limit $n=3$ in our final results exists and is
finite, predicting a larger unification scale, up to a factor of $\approx 10$
for unification at strong coupling together with a larger $\alpha_{3}(M_{Z})
$. However the higher order corrections are expected to be very large in
this case too so these results should be viewed with caution. This case has
been analysed in detail \cite{russell}, but  ignoring the gauge corrections
to thresholds which, as we have seen, affects the result considerably.

For $n=4$ or larger the perturbation expansion is well behaved for
intermediate values of the coupling. The results for the unification scale
and the strong coupling are shown in Figs 1 and 2.One may see that the
unification scale is increased, but by a factor $<5$ for $\alpha _{g}<0.45.$ 
The values of $\alpha _{3}(M_{z})$  always {\it increases} and are larger
than in MSSM. This is in disagreement with the analysis of \cite{hemp} due,
we think, to errors in the numerical intergration used.  In fact it can be
shown analytically that the partial derivative of $\alpha _{3}$ with respect
to the ratio $\alpha _{g}/\alpha _{g}^{o}$ is always positive for any fixed $%
n\ge 4$ with $\alpha _{g}\ge \alpha _{g}^{o}$ (see its expression given in
the Appendix). Therefore, in these cases, the strong coupling is larger then
in MSSM.  

The dependence of the  masses of the additional states on the unified
coupling is shown in Figs. 3, 4 and 5. One may see the masses increase as we
increase the number of states $n$, while keeping $\alpha _{g}$ fixed. Note
also that for a given $n$, there are two values of $\alpha _{g}$ which
predict the same value of $\mu _{q}$, which shows the advantage of
setting $\alpha _{g}$ as our input parameter instead of $\mu _{g}$, avoiding the
ambiguity of the methods having $\mu _{g}$ as an input parameter and $\alpha
_{g}$ as an output. This is particularly relevant when we consider the
predictions for large values of the unified coupling.   

To summarize, in contrast to previous analyses, we have found that the
addition of heavy states systematically increase $\alpha _{3}(M_{z})$ taking
it further from the current experimental value. The effect of the additional
states does increase the unification scale, but by a factor less than $%
\approx 5$ even for the largest couplings which may still be in the
perturbative domain. Since for unified couplings of $O(0.3)$ the string
prediction of eq(\ref{su}) increases by a factor $2.7$ there is only a
small improvement  with the weakly coupled heterotic string prediction.
Note that the analysis has neglected the Yukawa couplings of the
third generation which may be important; their effect is investigated 
elsewhere \cite{sun}.

\section{Massive threshold corrections}

The analytic results we obtained in Sections 5, 6 refer to the case when the
heavy multiplets have a common bare mass, $\mu _{g}$ at the unification
scale and are split only by gauge interactions. We found that the threshold
corrections introduced by this splitting largely cancel the two-loop
corrections in the beta functions leading to the form of eq(\ref{HMSSM}) or
equivalently eq(\ref{shif}). However Yukawa couplings will also cause the
massive spectrum to split introducing further threshold corrections.
Fortunately the form derived by Shifman, eq(\ref{shif}) shows that the
cancellation persists for Yukawa corrections as well and that the heavy
states contribute with their bare mass only; radiative corrections to the
masses should not be included if the appropriate two-loop beta function is
used. Thus the only effects of Yukawa couplings are contained in the $Z$
factors for the light fields. This is important because, unlike the
threshold effects, these corrections  are not proportional to $n$ 
and so we do not expect such effects to be large. 
There remains the corrections which may
occur in non-Grand-Unified models due to the possible splitting of the bare
Yukawa couplings  $\lambda _{\psi }$ introduced in Section 2. In this case
the correction factors to our previous results are given by

\begin{equation}
\frac{M_{g}}{M_{g}^{o}}=\left[ \frac{M_{g}}{M_{g}^{o}}\right] _{old}\left[ 
\frac{\mu _{d^{c}}^{o}}{\mu _{l}^{o}}\right] ^{\frac{n_{5}}{28}}\left[ \frac{%
{\mu _{e^{c}}^{o3}\mu _{u^{c}}^{o4}}}{{\mu _{q}^{o7}}}\right] ^{\frac{n_{10}%
}{28}}
\end{equation}
where the label ``old'' used in this section refers to our previously
obtained results of Sections 5 and 6 and $\mu _{\psi }^{o}$ stands for the
bare mass for $\psi $ field. Also: 
\begin{equation}
\alpha _{3}^{-1}(M_{z})-\alpha _{3}^{-1}(M_{z})_{old}=\frac{9n_{5}}{28\pi }%
\ln \left[ \frac{\mu _{l}^{o}}{\mu _{d^{c}}^{o}}\right] +\frac{3n_{10}}{%
28\pi }\ln \left[ \frac{\mu _{q}^{o7}}{{\mu _{u^{c}}^{o5}\mu _{e^{c}}^{o2}}}%
\right] 
\end{equation}
for the change in the strong coupling. The general effect of this
uncertainty requires a specific model for the bare couplings. However a
general feature emerges. For the case of additional $5s$ an increase in the
unification scale is always accompanied by an increase in the strong
coupling. For the case of additional $10s$ it is possible to avoid this, but
only through a very special choice of the bare masses which, lacking a
specefic model, looks quite unnatural. Further one may see that the change in
the unification scale is quite small for reasonable numbers of additional
multiplets unless the bare masses are split by many orders of magnitude. In
string theories the differences between the bare Yukawa couplings are typically
of $O(1)$ for the allowed couplings so it seems unlikely such bare threshold
corrections will give a large change to the unificaion scale.

\section{Summary}

Motivated by the structure of many Grand Unified and Superstring theories,
we have considered the implications of extending the structure of the MSSM
by the addition of extra copies of massive vectorlike $(I+\bar{I})$
representations. These we take to fill out complete $SU(5)$ multiplets, even
though there may be no stage of Grand Unification. While such
representations do not affect the one loop unification predictions, they can
have a substantial effect through their threshold effects and at the
two-loop level. We computed their effects on the unification predictions for
the running gauge couplings for the case the additional representations have
common bare mass. We found that they always {\it increase} the value 
of $\alpha _{3}$ (ignoring Yukawa couplings effects) and thus, cannot 
explain the (marginal) discrepancy between the
predicted value $\alpha _{3}(M_{Z})\ge 0.125$ and the observed value 
$\alpha_{3}(M_{Z})=0.118\pm 0.003$.

We also found that the unification scale is systematically increased by the
addition of such heavy states. The increase depends on the value of the
coupling at unification and is largest for the case this is approaching the
strong coupling limit. For this case the unification scale increases by a
factor varying (function of $n$) between 1 and $5$ for 
$\alpha_{g}$=$0.3,$ at the limit of the perturbative analysis. The
larger enhancement, allowing for the change in the string prediction at
larger coupling,  is only slightly closer to the value predicted in the
(weakly coupled) heterotic string theory. 

\section{Acknowledgments}
The authors want to  thank G. Amelino-Camelia, K. Dienes, D.R.T. Jones, 
I. Kogan, J. March-Russell, M. Shifman for useful discussions.
D.G. gratefully acknowledges the financial support from the part of
University of Oxford and Oriel College (University of Oxford).

\section{Appendix}
\appendix
The one loop beta function in the presence of complete $SU(5)$ multiplets is
given by:

\begin{equation}
b_i^{\prime}=\left( 
\begin{array}{r}
\frac{33}{5}+n \\ 
1+n \\ 
-3+n
\end{array}
\right)
\end{equation}
where $n$ represents the linear combination $n=(n_5+3n_{10})/2$, with $%
n_5=N_{{\bf 5}}+N_{{\bf \overline{5}}}$ and $n_{10}=N_{{\bf {10}}}+N_{{\bf 
\overline{10}}}$. The one loop MSSM beta function is given by $%
b_i=b^{\prime}_i(n=0)$. \newline
The two loop beta function in the presence of the complete $SU(5)$
multiplets is given by:

\begin{equation}
b_{ij}^{\prime }=\left( 
\begin{array}{rcc}
\frac{199}{25}+\frac{7}{30}n_{5}+\frac{23}{10}n_{10} & \frac{27}{5}+\frac{9}{%
10}n_{5}+\frac{3}{10}n_{10} & \frac{88}{5}+\frac{16}{15}n_{5}+\frac{24}{5}%
n_{10} \\ 
&  &  \\ 
\frac{9}{5}+\frac{3}{10}n_{5}+\frac{1}{10}n_{10} & 25+\frac{7}{2}n_{5}+\frac{%
21}{2}n_{10} & 24+8n_{10} \\ 
&  &  \\ 
\frac{11}{5}+\frac{2}{15}n_{5}+\frac{3}{5}n_{10} & 9+3n_{10} & 14+\frac{17}{3%
}n_{5}+17n_{10}
\end{array}
\right) 
\end{equation}
The MSSM two loop beta function is given by $b_{ij}=b_{ij}^{\prime
}(n_{5}=0,n_{10}=0)$. \newline
The values of the coefficients $\beta _{j,\psi }$ used in text, are given by
the following matrix with the symmetry group index j as a line index $%
j=\{1,2,3\}$, and with the type of the field $\psi $ as a column index
running over the set $\{l,d^{c},e^{c},u^{c},q\}$, in this order. 
\begin{equation}
\beta _{j,\psi }=\left( 
\begin{array}{ccccc}
\frac{3}{10} & \frac{1}{5} & \frac{3}{5} & \frac{4}{5} & \frac{1}{10} \\ 
\frac{1}{2} & 0 & 0 & 0 & \frac{3}{2} \\ 
0 & \frac{1}{2} & 0 & \frac{1}{2} & 1
\end{array}
\right) _{j,\psi }
\end{equation}
In the same notation we have for the matrix $\eta _{j,\psi }$ 
\begin{equation}
\eta _{j,\psi }=\left( 
\begin{array}{ccccc}
\frac{3}{10} & \frac{2}{15} & \frac{6}{5} & \frac{8}{15} & \frac{1}{30} \\ 
\frac{3}{2} & 0 & 0 & 0 & \frac{3}{2} \\ 
0 & \frac{8}{3} & 0 & \frac{8}{3} & \frac{8}{3}
\end{array}
\right) _{j,\psi }\equiv 2C_{j}(\psi )
\end{equation}
where $C_{j}(\psi )$ is the quadratic Casimir operator, given by $3Y_{\psi
}^{2}/5$, 3/4, 4/3 for $U(1)$, $SU(2)$, $SU(3)$ respectively. \newline
With these definitions, we get, 
\begin{equation}
\sum_{\psi =\{l,d^{c},e^{c},u^{c},q\}}N_{\psi }\,\beta _{i,\psi }\,\eta
_{j,\psi }=\frac{1}{2}\left( b_{ij}^{\prime }-b_{ij}-2n\lambda _{j}\delta
_{ij}\right) 
\end{equation}
with $\lambda _{1}=0\,$, $\lambda _{2}=2\,$, $\lambda _{3}=3$.\newline
Throughout the text we used $SU(5)$ normalization for gauge couplings, $%
t=1/(2\pi )\ln (scale/M_{g})$, $\alpha _{i}=g_{i}^{2}/(4\pi )$, $Y_{\psi
}=\lambda _{\psi }^{2}/(4\pi )$.\newline
In two loop, the equation for $\xi $ used in Section 4 is 
\begin{equation}
\alpha _{g}^{-1}-\alpha _{g}^{o-1}+n\xi =-\frac{1168}{231\pi }\ln \left[ 
\frac{\alpha _{g}}{\alpha _{g}^{o}}\right] +\sum_{j=1}^{3}\Delta _{j}\ln
\left[ 1+b_{j}^{\prime }\alpha _{g}\xi \right] 
\end{equation}
with $\Delta _{j}$ given by 
\begin{equation}
\Delta _{j}=\left\{ \frac{13n}{1320b_{1}^{\prime }\pi },\frac{291n}{%
56b_{2}^{\prime }\pi },\frac{-24n}{7b_{3}^{\prime }\pi }\right\} 
\end{equation}
The derivative of $\alpha _{3}(M_{z})$ (eq.(\ref{res4})), with
respect to $z=\alpha _{g}/\alpha
_{g}^{o}$ is positive, given that $z\ge 1$ and $n\ge 4$ as it can be seen
from its expression given below: 
\[
\frac{\partial }{\partial z}\alpha _{3}(M_{z})=3\,\left[ \,320+240\,\left(
-4+n\right) +60\,{{\left( -4+n\right) }^{2}}+5\,{{\left( -4+n\right) }^{3}}
+2720\,\left( -1+z\right) \right. 
\]
\[
+1600\,\left( -4+n\right) \,\left( -1+z\right) +290\,{{\left( -4+n\right) }
^{2}}\,\left( -1+z\right) +15\,{{\left( -4+n\right) }^{3}}\,\left(
-1+z\right) 
\]
\[
+5716\,{{\left( -1+z\right) }^{2}}+2789\,\left( -4+n\right) \,{{\left(
-1+z\right) }^{2}}+400\,{{\left( -4+n\right) }^{2}}\,{{\left( -1+z\right) }
^{2}}
\]
\[
+15\,{{\left( -4+n\right) }^{3}}\,{{\left( -1+z\right) }^{2}}+496\,{{\left(
-1+z\right) }^{3}}+1429\,\left( -4+n\right) \,{{\left( -1+z\right) }^{3}}
\]
\[
\left. +170\,{{\left( -4+n\right) }^{2}}\,{{\left( -1+z\right) }^{3}}+5\,{{
\left( -4+n\right) }^{3}}\,{{\left( -1+z\right) }^{3}}\right] \times 
\]
\[
\left\{ {14\,\pi \,\left[ n+5\,\left( -1+z\right) +\left( -4+n\right)
\,\left( -1+z\right) \right] }\times \right. 
\]
\[
\left[ 20+5\,\left( -4+n\right) +53\,\left( -1+z\right) +5\,\left(
-4+n\right) \,\left( -1+z\right) \right] \times 
\]
\begin{equation}
\left. z\,\left[ -1+n+\left( -4+n\right) \,\left( -1+z\right) +z\right]
\right\} ^{-1}\alpha _{3}^{2}(M_{z})
\end{equation}

\newpage\noindent {\Large Figure captions} \vskip 1cm \noindent Figure 1.
The ratio $M_g/M_g^{o}$ plotted in function of the ratio $\alpha_g/\alpha_g^o
$ for different values of $n$. \vskip 5mm \noindent Figure 2. The values of $%
\alpha_3(M_z)$ plotted in function of the ratio $\alpha_g/\alpha_g^o$ for
different values of $n$. \vskip 5mm \noindent Figure 3. The values of $%
\log_{10}(\mu_{e^c}/M_g^o)$ plotted in function of the ratio $%
\alpha_g/\alpha_g^o$ for different values of $n$. \vskip 5mm \noindent
Figure 4. The values of $\log_{10}(\mu_{q}/M_g^o)$ plotted in function of
the ratio $\alpha_g/\alpha_g^o$ for different values of $n$. \vskip 5mm
\noindent Figure 5. The values of the splitting $\log_{10}(\mu_{q}/\mu_{e^c})
$ plotted in function of the ratio $\alpha_g/\alpha_g^o$ for different
values of $n$. \newpage %
\psfig{figure=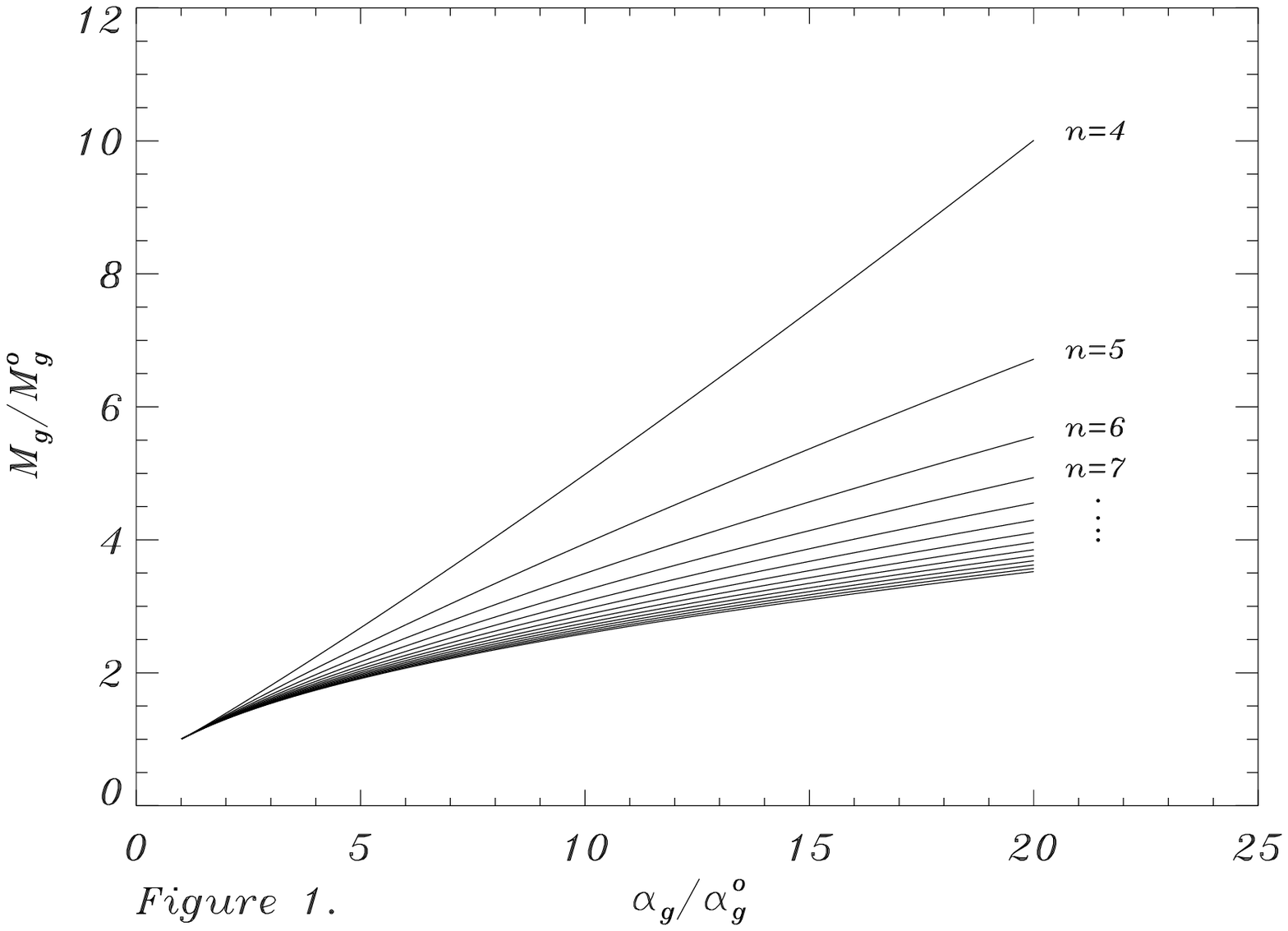,height=11cm,width=13cm} %
\vskip 0.5cm %
\psfig{figure=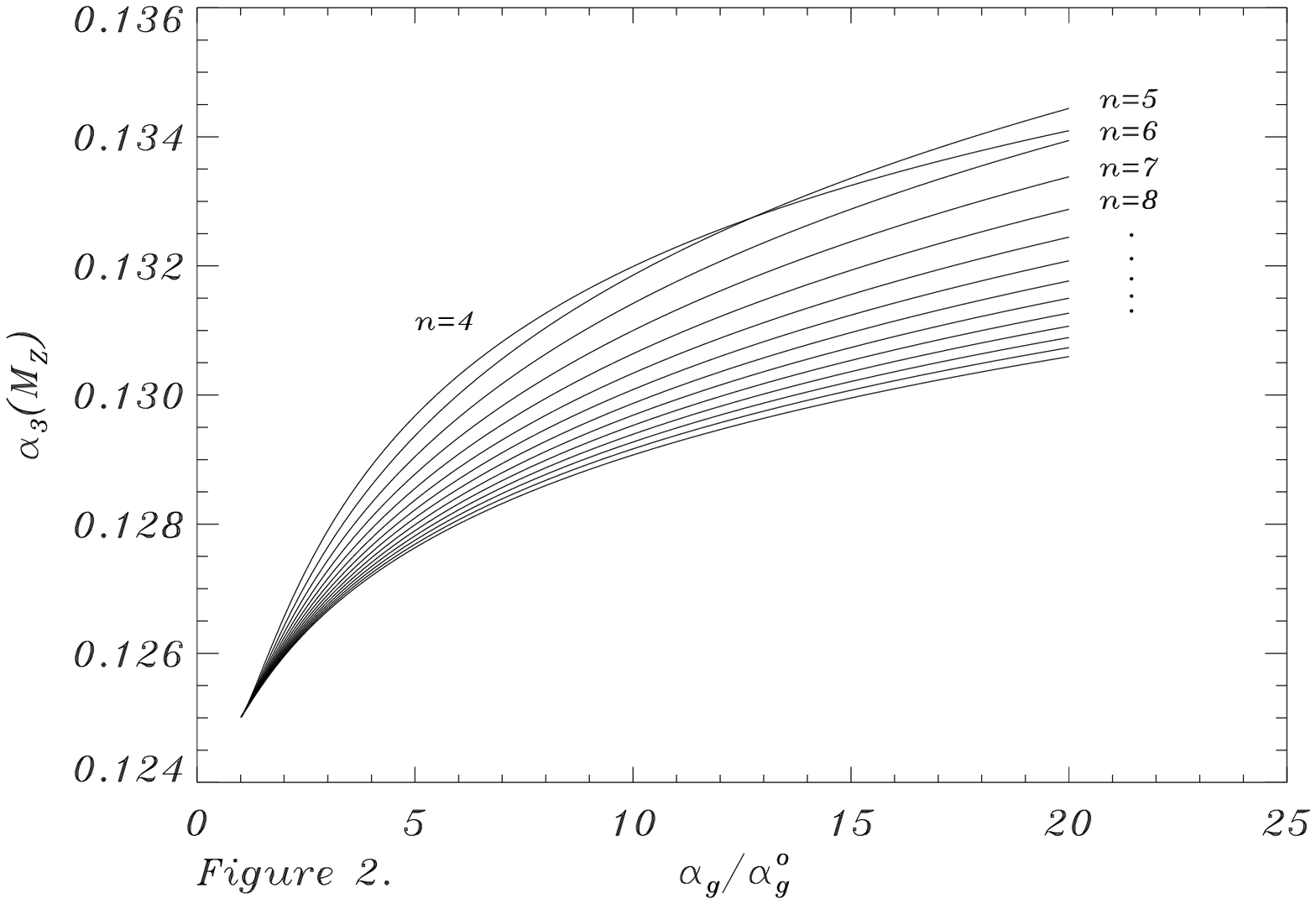,height=11cm,width=13cm}
\newpage %
\psfig{figure=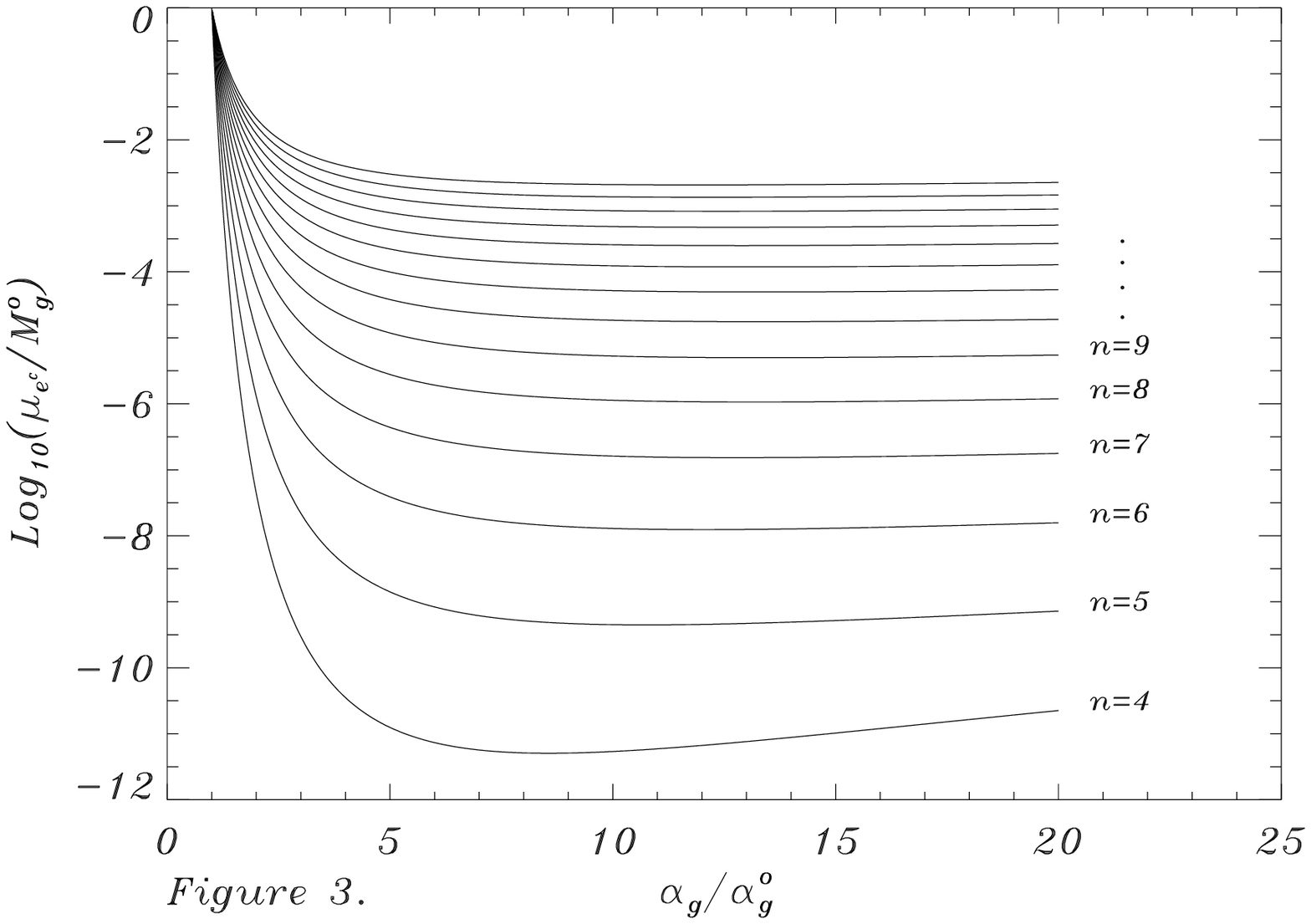,height=11cm,width=13cm} %
\vskip 0.5cm %
\psfig{figure=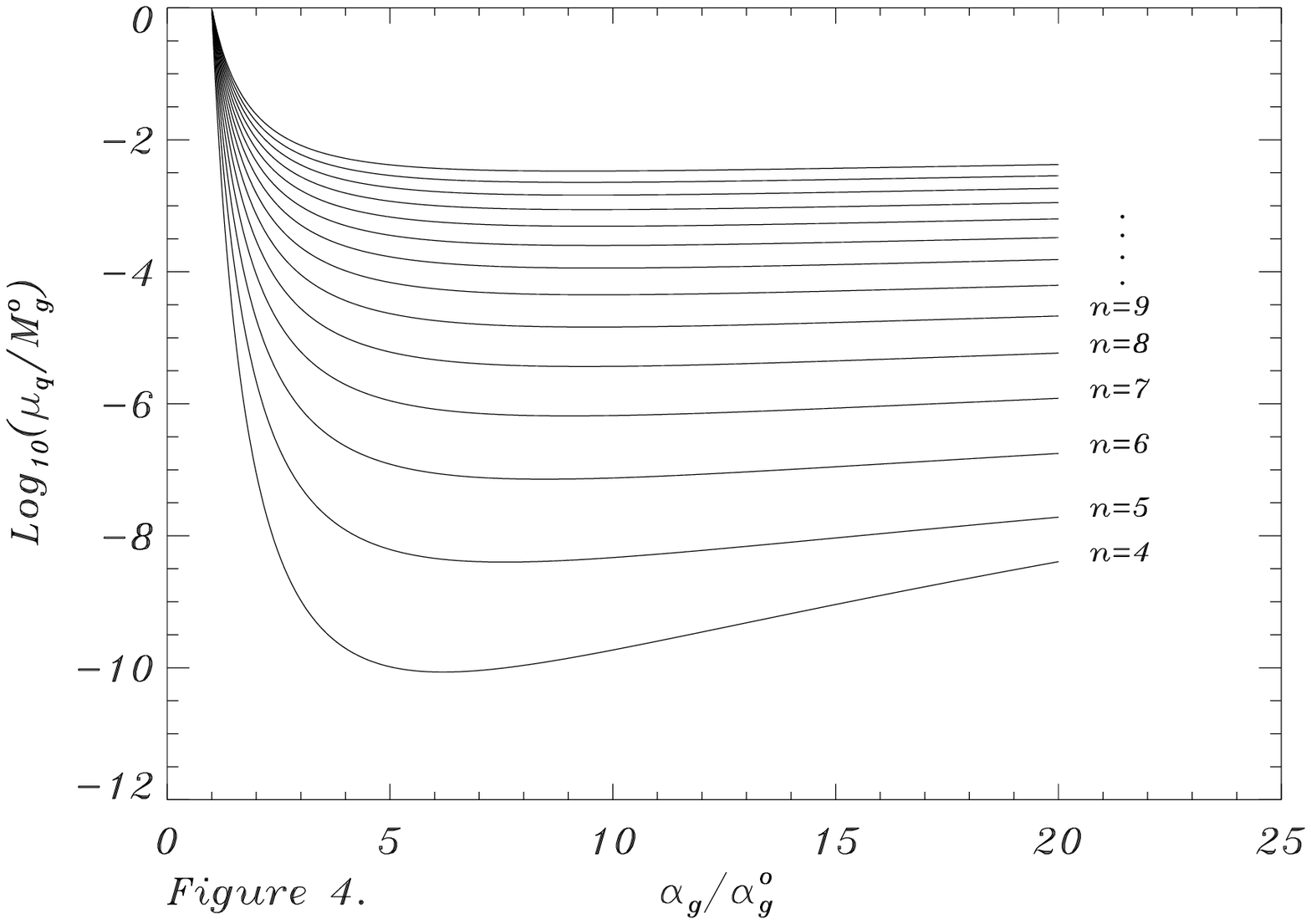,height=11cm,width=13cm}
\newpage %
\psfig{figure=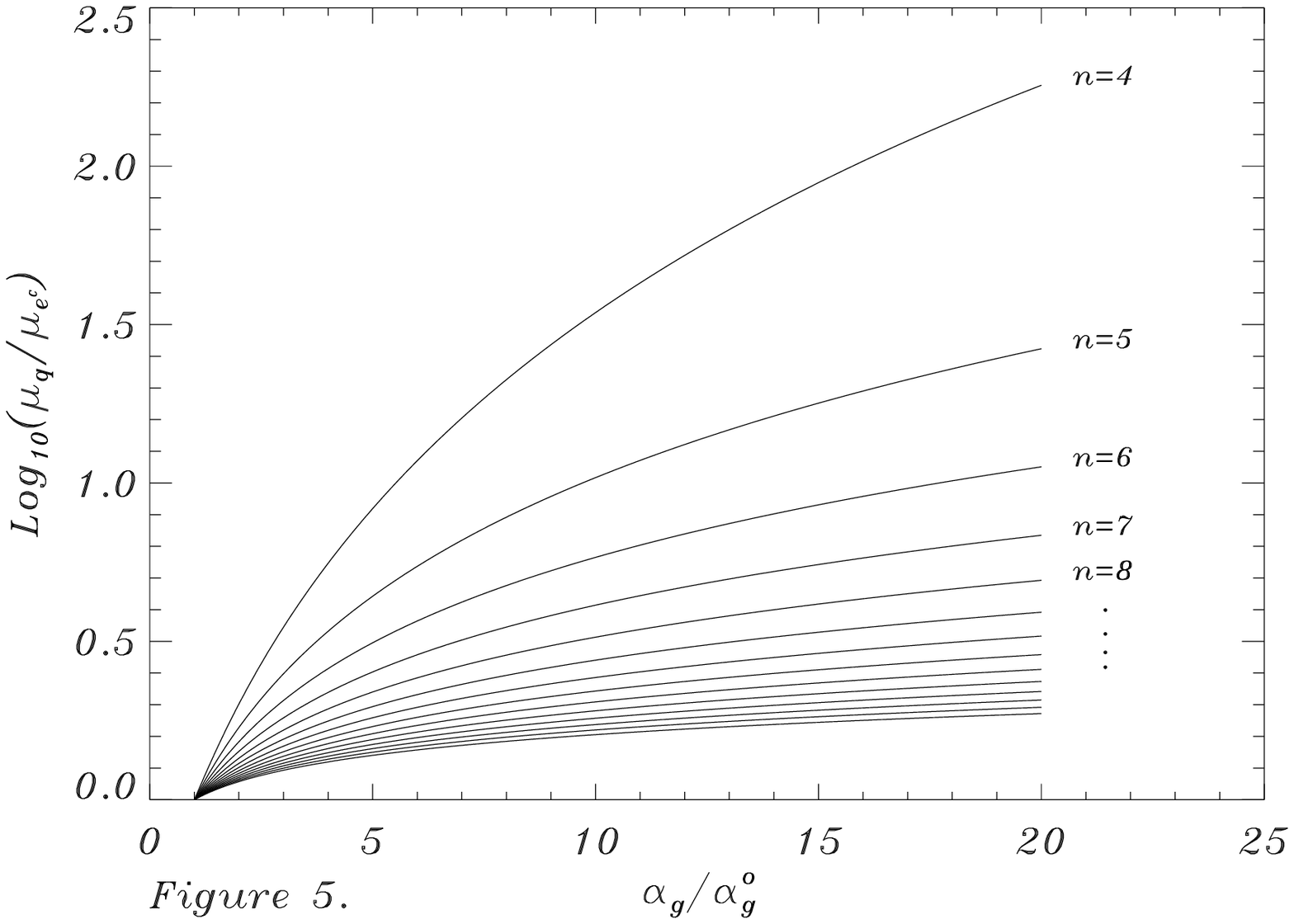,height=11cm,width=13cm}

\end{document}